# Spoken Humanoid Embodied Conversational Agents in Mobile Serious Games: A Usability Assessment

*Danai Korre and Judy Robertson*
*The University of Edinburgh, Edinburgh, UK*

**ABSTRACT**

This paper presents an empirical investigation of the extent to which spoken Humanoid Embodied Conversational Agents (HECAs) can foster usability in mobile serious game (MSG) applications. The aim of the research is to assess the impact of multiple agents and illusion of humanness on the quality of the interaction. The experiment investigates two styles of agent presentation: an agent of high human-likeness (HECA) and an agent of low human-likeness (text). The purpose of the experiment is to assess whether and how agents of high human-likeness can evoke the illusion of humanness and affect usability. Agents of high human-likeness were designed by following the ECA design model that is a proposed guide for ECA development. The results of the experiment with 90 participants show that users prefer to interact with the HECAs. The difference between the two versions is statistically significant with a large effect size (d=1.01), with many of the participants justifying their choice by saying that the human-like characteristics of the HECA made the version more appealing. This research provides key information on the potential effect of HECAs on serious games, which can provide insight into the design of future mobile serious games.

## 2. INTRODUCTION

The latest generation of mobile devices has the capabilities of supporting more complex applications in terms of technical and interactive features. The portability and wireless access to the internet makes mobile devices a tool of great potential for formal and informal edification. However, there is a lack of studies regarding the use and the effectiveness of mobile devices for this purpose (Y.-T. Sung et al., 2016).The multi-touch nature of the mobile interaction along with the smaller screen size and the human fingertip call for a more compact information architecture (IA) with cleaner user interfaces and a smaller number of steps (Doumanis et al., 2015).  The way users interact with mobile devices is changing again since the latest generation of mobile devices includes voice-driven virtual assistants (Siri, Google now, S voice) (Santos-Perez et al., 2013).

Embodied conversational agents (ECAs) are virtual characters with the ability to converse with a human through verbal (speech) and/or non-verbal communication (text and/or gestures) (Cassel et al., 2000). There are many theoretical advantages in favour of ECAs and spoken dialogue systems (systems that use speech as input) and it is assumed they provide a more "natural interaction" (Weiss et al., 2015, Takeuchi and Naito 1995). They are often considered anthropomorphic entities due to the linguistic, extra-linguistic and non-verbal information they convey. The anthropomorphisation of interfaces evokes an illusion of humanness from the user's behalf that can affect the interaction and subsequently the usability.

Increased believability and perceived trustworthiness are a major goal in ECA research. To achieve that, human-like virtual agents are often developed; this human-like aspect makes ECAs subject to social conventions (Gris Sepulveda, 2015). According to some studies, the interaction with spoken dialogue systems, either in the form of an embodied agent or not, is still inferior compared to other approaches that allow a direct manipulation of the system to which the user responds instinctively, despite the theoretical advances of ECAs and dialogue systems (Weiss et al., 2015). However, the year 2016 has been a tipping point for conversational interfaces with major companies investing heavily in this technology (McTear, 2017). Furthermore, conversational agents are expected to be important modes of interaction in Virtual Reality (VR) environments (Beilby and Zakos, 2014) Thus, the importance of understanding the quality attributes and the issues that are affiliated with the development of high-quality and usable conversational agents, increases exponentially (Radziwill and Morgan, 2017).

The introduction of an ECA without taking into consideration the context of use and the purpose of the system could lead to a poor performance by the user as the ECA might act as a distraction rather than a helpful element and the interaction may be frustrating for the user (Doumanis et al., 2015). Therefore, whether usability and quality are to be enhanced by using an ECA in a multimodal human-machine interface must be decided for each application anew (Weiss et al., 2015). This is also a strong reason to examine if and how ECAs enhance usability over current interaction paradigms in serious game (SG) environments, even more so in mobile devices as there is a recent trend towards mobile serious games (MSG) and empirical evidence is limited (Gamelearn, 2015; Adkins, 2020; Doumanis, 2015).

Numerous aspects of ECAs (physical, behavioural etc.) have been evaluated empirically for several years. ECAs' interdisciplinary nature allows for further investigation on how they can create highly usable interfaces, as they rely heavily on technological advances such as the processing power, rendering techniques, graphic cards that are ever changing thus making previous research dated or even obsolete. As technology and computational techniques advance, ECAs remain a thriving research topic; in the last year with the HCI community new ECA developments across contrasting domains were published – for example an ECA tutor for teaching fractions to children (Krishna, Pelachaud, & Kappas, 2020), an animated daisy plant companion for older adults (Simpson, Gaiser, Mac'ik, & Bressgott, 2020), a virtual agent designed to increase productivity at work (Grover, Rowan, Suh, McDuff, & Czerwinski, 2020) and an ECA to assist in the area of substance use counselling (Olafsson, Wallace, & Bickmore, 2020).

Although the evidence for the adoption of ECAs is encouraging, much experimental work related to the media equation and ECAs has been conducted on desktop computers. Mobile users may have a different reaction towards ECAs as mobile devices are most commonly used in places with ambient noise and crowds. Another issue is that research on mobile ECAs dates to early 2000 while the mobile and computer graphics technology has seen tremendous changes in recent years and most users are more technology literate. Based on these observations, further research on mobile ECAs is necessary. This paper presents an empirical investigation of the extent to which spoken Humanoid Embodied Conversational Agents (HECAs) can foster usability in mobile serious game (MSG) applications. The aim of the research is to empirically assess the impact of multiple agents and illusion of humanness on the quality of the interaction.

The rest of the paper is organized as follows: in the background section, a literature review is given which includes theories on the use of ECAs, the proposal of the ECADM model for categorisation of ECAs and the experimental interface design. The evaluation section reports results of a qualitative and quantitative analysis of a user study with 90 participants which compares two styles of agent presentation: an agent of high human-likeness (HECA) and an agent of low human-likeness (text). The results of the study are then discussed, with consideration of future work and implications for developers.

## 2.1. Background

### 2.1.1. Embodied conversational agents (ECAs)

The term "Embodied Conversational Agent" was coined by Justine Cassell in 2000 and is defined as follows: "computer interfaces that can hold up their end of the conversation, interfaces that realise conversational behaviours as a function of the demands of dialogue and as a function of emotion, personality, and social conversation" (Cassell et al., 2000). According to Cassell, these embodied conversational agents (ECAs) are virtual characters with the ability to converse with a human through verbal (speech) and/or non-verbal communication (text and/or gestures). Interface agents such as ECAs are agents that have some form of a graphical/visual representation on the interface and are capable of autonomous actions without explicit directions from the user (Doumanis and Smith, 2015). The terms mostly used interchangeably with ECAs are: virtual character, intelligent agent or social agent (Veletsianos and Miller, 2008).

Another term related to ECAs is that of virtual humans. Virtual humans are the result of the emergence of different fields around computer science such as artificial intelligence, computer animation, computer graphics, human-computer interaction and cognitive science (Kasap and Magnenat-Thalmann, 2008). These characters can play the role of the guide, the trainer, the teammate, the rival or a source of motion in virtual space (Brogan et al., 1998). However, virtual humans along with their complexity, can vary diametrically as each of them has a specific role and purpose depending on the goal of the application. The main difference between ECAs and virtual humans is that virtual humans always have the appearance of a human and they do not necessarily possess any intelligence or communication skills. An example is the non-interactive characters in games that are used to populate a scene. When virtual humans are combined with ECAs, the result is a Humanoid Embodied Conversational Agent (HECA).

Embodied conversational agents need to possess the following abilities which comply to the modelling of regular autonomous agents. First, they should perceive verbal and/or nonverbal input from the user and the user's environment. Second, they should translate the inputs' meaning and respond appropriately through verbal and/or nonverbal actions. Last, those actions should be performed by an animated computer character in a virtual environment (Huang, 2018).

According to De Vos, (2002), ECAs share the following features: anthropomorphic appearance (human, animal or fantasy figure); a virtual body that is used for communication purposes; natural communication protocols; multimodality and performing a social role.

This last feature is of particular interest for the research reported in this paper. Embodied conversational agents are different from other computer systems in the sense that they try to emulate human-to-human interaction in a believable manner and, therefore, have a social standing. The concept of believability is described by Bates, (1994) as "one that provides the illusion of life, and thus permits the audience 's suspension of disbelief". In ECA research, the concept of believability is approached in two ways. One way is that higher believability can result by implementing more NL functions (Cassell and Stone, 1999). The other way is that believability is more a matter of personality and emotions supported by the significant roles that portayal of emotions plays in creating "believable" characters by Disney (Bates, 1994). The work presented in this paper uses ECAs that express personality and emotions as part of the "illusion of humanness" of the system.

### 2.1.2. The illusion of humanness

One of the major theoretical foundations of virtual character and ECA research is the media equation. Nass, et al. (1994) proposed the "Computers as Social Actors (CASA)" approach that is now known as the media equation theory. It implies that people tend to interact with computers and media in an inherently social way. Even though the users know that the computer is a medium rather than a human being, they treat it in a social way as they would in human-human interaction (Nishida et al., 2014).

Experimental demonstration of this effect was carried out by Reeves and Nass (1996) showing that humans treated computers and media in an inherently social way although not consciously. The users rated seemingly "polite" computers as more favourable even though computers are not capable of expressing politeness. As a result, human-like interfaces such as virtual agents, pedagogical agents and ECAs would also be in principle subjected to social rules (Veletsianos, 2010). According to Kramer et al., (2015) the effects of ECAs can be described as "social" if they can evoke to the participant similar emotional, cognitive or behavioural reactions to the ones evoked by other humans.

Further research by Nass and colleagues (Nass et al., 1997; Nass and Moon, 2000) used the term "Ethopoeia" to describe the phenomenon that occurs during the interaction between a human and a virtual agent. The "Ethopoeia" explanation suggests that people unconsciously apply social rules when interacting with a virtual agent in a similar way they would with other humans. Additionally, they reject the hypothesis that people consciously anthropomorphised computers thus they replied consciously as participants, but when asked denied doing so. The explanation according to Nass and colleagues can be found in the way the human brain has evolved to automatically recognise emotive reactions from humans (Kramer et al., 2015). Studies supporting this notion have provided evidence that users/participants ethnically identify with virtual agents, respond politely and apply gender stereotypes to them (Scott et al., 2015).

For the purposes of this research the illusion of humanness is defined as *the user's perception that the system possesses human attributes and/or cognitive functions*. The illusion of humanness is not to be confused with anthropomorphism which is more related with the attribution of human properties to non-human entities or humanoid which almost always refers to having the appearance of a human. When it comes to anthropomorphism, "attribution" is a key term as it implies that giving human characteristics to non-human agents is a conscious action from humans' side while "the illusion of humanness" is an involuntary reaction to a humanoid and anthropomorphic interface. The illusion of humanness is an extension of the "ethopoeia" explanation and persona effect but not limited to the unconscious application of social rules or an affective impact on learning but rather a determining factor on users' performance and perceived usability. It refers more specifically to systems that present information by utilising one or more human-like attributes (ex. voice, gaze, gestures, body) thus giving an illusion of "humanness" to the user. These attributes can be presented in textual, auditory and/or visual form. These attributes can be in the form of: gesturing, facial expression, eye gaze, human-like movement, voice, embodiment and behaviour (ex. using pronouns, personality, politeness, humour).

The illusion of humanness is related to anthropomorphism but is not synonymous. Anthropomorphism is the attribution of human characteristics to non-human entities and is a combination of the Greek words for human and form/appearance (ἄνθρωπος + μορφή). The word "anthropomorphism" etymologically is more relevant to the appearance, but it has also been used in the past to describe human-like behaviour in the field of HCI or even an umbrella term for human-like interfaces therefore it will be briefly explored as the factor the evokes the "illusion of humanness" effect.

The psychology of anthropomorphism was examined by Adam Waytz (Harvard University) and Nicholas Epley (University of Chicago). This neuroscience research revealed that when people think of humans and non-human entities, the same brain areas are activated. This result is an indication that anthropomorphism utilises the same processes as the ones used when thinking of other people. Thus, anthropomorphism can evoke a certain mental response (illusion of humanness) where people think of non-human entities as human consequently render them worthy of consideration or moral care (Waytz et al., 2014).

Anthropomorphism can take on various forms at the user interface. The simplest form is *textual*, another form concern using *auditory cues* while *visual cues* of multiple manifestations can be used and typically would involve using text and/or voice audio (Murano, 2006) (Table 1). The anthropomorphised aspect of textual feedback is the way text is written on the screen, i.e. using pronouns such as "I". Some chatbots are also an example of displaying textual anthropomorphism or personification. Auditory cues or "voice" are usually expressed in the form of Text-to-speech (TTS) technology or dynamically loaded voice clips of humans. The system may also use pronouns such as "I" to refer to itself. An example of a system using auditory cue is the virtual assistants that have recently became popular. Home virtual assistants such as Amazon Alexa and Google home but also mobile virtual assistants such as Siri, S voice and ok Google are a few examples of virtual assistants that speech recognition and voice output of a TTS form. Some of these systems have names associated with them such as Alexa and Siri which give the illusion of an identity and further anthropomorphises the system.

The mere existence of voice expresses anthropomorphism. Even for the systems with no allocated names, the mere fact that they have a human-like voice gives an illusion of persona or identity due to extra-linguistic data provided through the voice beyond the context of the message such as intonation, gender etc. When humans hear a voice, they can in most cases understand emotion based on the tone that is being used. Speech can reveal cues to the speaker's personality, beliefs, cognitive process, social membership etc. (Zara, 2007).

Non-verbal communication and extra-linguistic information are also of importance and can be anthropomorphic. Developing ECAs that mimic humanlike non-verbal behaviours reinforces the understanding that the inclusion of non-verbal behaviour enhances the human-agent interaction. Images that are characterised as anthropomorphic can range from simple stick drawings to hyper realistic 3D characters (Murano, 2005). This includes video clips of humans (Bengtsson et al., 1999). Non-verbal behaviour includes but is not limited to lip-synching that is accurate with ECA speech output, gesturing, facial animations such as eyebrow raising and change of eye gaze. Face animations (rising of eyebrow, smiling etc.) have been used successfully to communicate emotion and signal speech input from the user (Doolin, 2014). Through a series of experiments Foster (2007) found that when speech is combined with appropriate hand gestures, the usability of human-ECA interaction is significantly enhanced.

| **Textual cues** | **Auditory cues** | **Visual cues** |
|---|---|---|
| Example: Virtual companion chatbots that use pronouns or names such as Replika.[1] | Example: Home virtual assistants that use pronouns or names such as Amazon's Alexa. | Example: Telespin's CoPilot soft skills training platform with the virtual reality firing training module (Barry).[2] |

**Table 1 - Forms of anthropomorphism with examples**

## 2.2. The ECADM Model for categorizing ECAs

The ECA Design Model model (ECADM) which organises ECAs' characteristics into three categories, is shown in fig. 1. This model serves a dual function: 1) inform design decisions for designers and 2) act as a guide to categorise ECA research which will allow for better comparisons and analyses.

On the presentation level, ECAs can be depicted as either human or non-human characters, animated or static, photorealistic or more stylised, 2D or 3D, they can have a full body, only a head, a bust or a torso and finally their physical properties can vary (hair colour, clothes, body type, accessories, age etc.) (Haake and Gulz, 2009; Gulz and Haake, 2006; Veletsianos and

---

[1] Found in: https://replika.ai/
[2] Found in: https://www.talespin.company/copilot

Miller, 2008; Clarebout and Heidig (née Domagk), 2012). Secondly on the interaction level, decisions on the input and output modalities of the ECA must be taken. Multimodality is a basic feature of ECAs; this means that ECAs can employ one or more of the inputs and output modalities such as voice and text.

Finally, the persona level of the ECA is constituted by features related to the perceived by the user character of the ECA. Just like in real life as well as with virtual assistants, voice plays a major role in forming opinions about someone's personality. The agent's voice along with their role in the application and the personality they adopt form a cluster of personality pointers. These personality pointers are also informed by non-verbal and extra-linguistic information.

Those categories are general and can be broken down to specifics, for example under the Interaction level one may add the number of agents within the application.

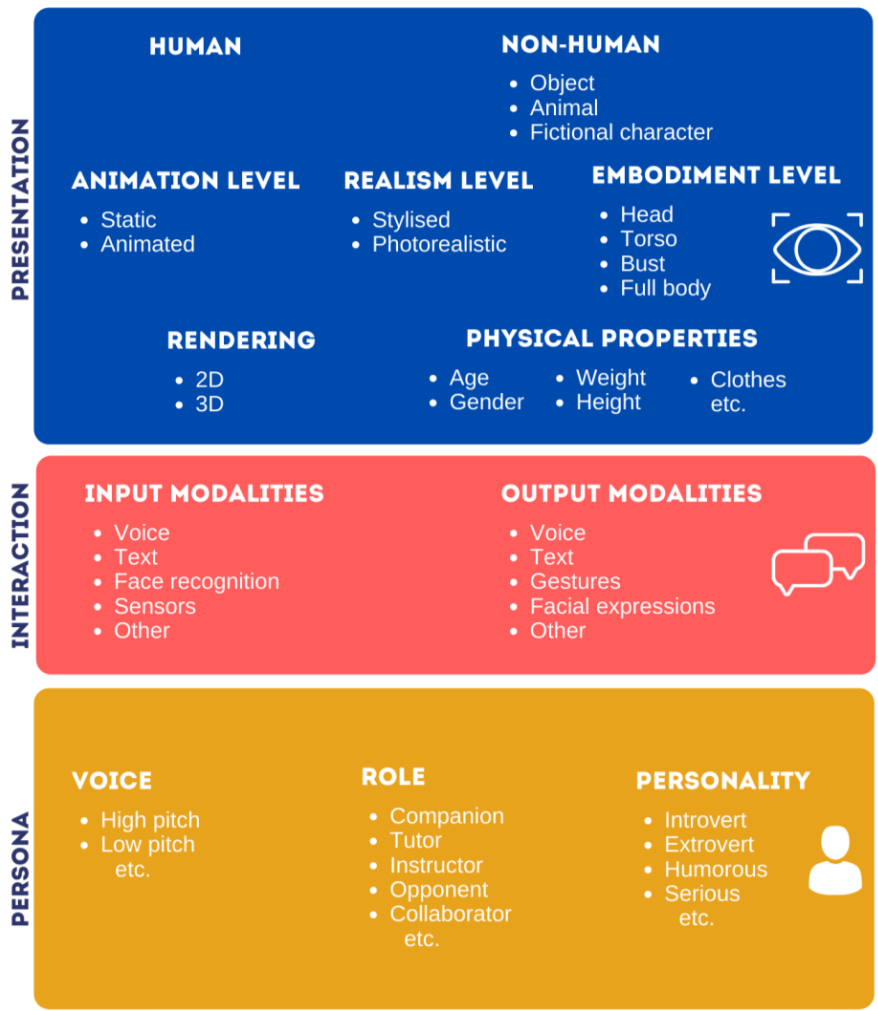

**Figure 1. Categories of ECA Design Model (ECADM).**

# 3. EXPERIMENTAL INTERFACE DESIGN

A system called Moneyworld was developed to examine the impact of multiple agents and illusion of humanness on the quality of the interaction for the user.

Isbister and Doyle, (2002) claim that an agent with physical appearance, sound and animation can cause a powerful visceral reaction on the user – evoke the "illusion of life". By enhancing realism in movement, creating natural sounding speech and creating the right visual style that fits the application, user's reaction to the agent can be amplified. Based on the assumptions that human-like realism can evoke an illusion of life and subsequently an illusion of humanness, two versions of agent representation are put to the test based on the spectrum of application interface design in relation to human likeness (Figure 2)

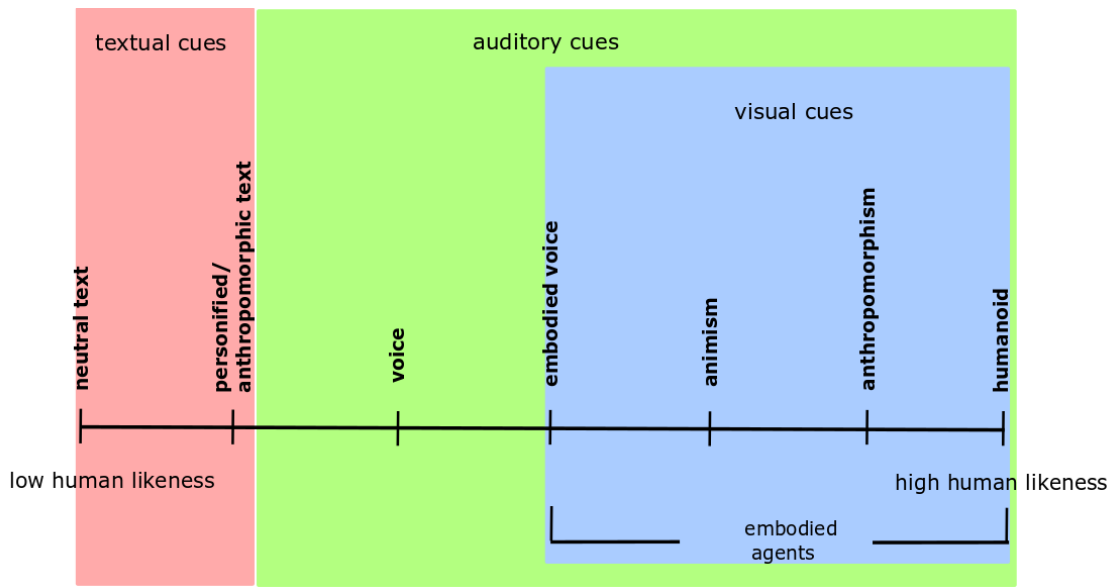

**Figure 2. Spectrum of application interface design in relation to human likeness.**

In order to achieve high-human likeness, a series of design decisions were made by following the ECADM (Figure 3). The choices were based on the literature which suggested that realism in all levels evokes the illusion of life. For the purposes of this research, two versions of a finance-related SG were compared, the high human-likeness version where the agents were represented by a humanoid ECA and a low human-likeness version where the agents are represented by neutral text conversational agents. Two agents for distinctive purposes (collaborator and instructor) were chosen to explore the dimension of the role of the agent. A mobile serious game was chosen as the application area because mobile gaming is expected to represent more than

50% of the total games market in 2020[3], there is a significant trend towards mobile serious games (Adkins, 2020) and empirical evaluations in mobile SGs are limited (Jordine et al., 2016). Serious games are defined as games, therefore interactive, with a clear goal, based on a set of rules and provide feedback (Wouters, 2013); whose primary purpose is not entertainment or enjoyment (Michael and Chen, 2005); yet they are fun to play and/or engaging, have a scoring system (feedback) and teach a skill, knowledge or attitude to the user that can be then used in the real world (goal) (Bergeron, 2006); and "a mental contest, played with a computer (interactive) in accordance with specific rules (rules) that uses entertainment to further government or corporate training, education, health, public policy, and strategic communication objectives" (Zyda, 2005).

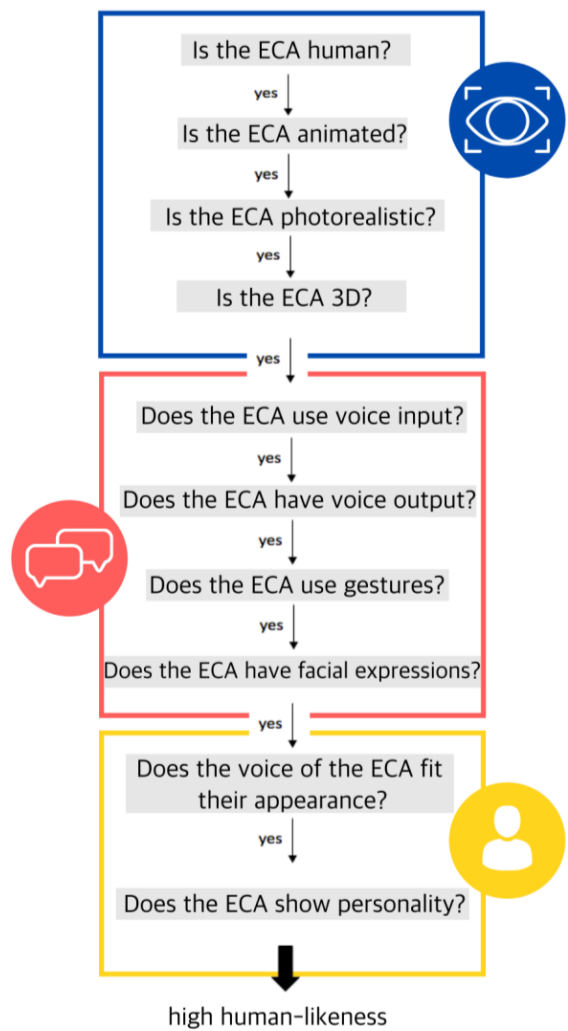

**Figure 3. ECA design decisions that result in high human-likeness**

---

[3] Reported by Newzoo: https://newzoo.com/insights/articles/the-global-games-market-will-reach-108-9-billion-in-2017-with-mobile-taking-42/

Computer games are undisputedly popular in modern society. Statistics show that the games industry is the fastest growing entertainment industry with 2.2 billion people playing games around the world. This fast growth is attributed to the popularity of games especially among younger people making them a great medium to obtain information and knowledge (Lenhart et al., 2008; Seng and Yatim, 2014; Korre, 2012). Mondly, a company that makes language learning applications claim that "The new generation of learning should be about gamified, immersive experiences that always make the users crave for more." (Rajnerowicz, 2020).

Moneyworld is a 3D interactive mobile serious game where the user travels back in time in order to learn more about the old money system that was used in the UK till the early 1970s. In this application, two photorealistic agents equipped with speech recognition are used. The participant partakes in a shopping experience using voice and mouse as input methods.

In the game introduction a female unembodied voice welcomes the user to the time machine chamber and introduces the concept of the application (Figure 4).

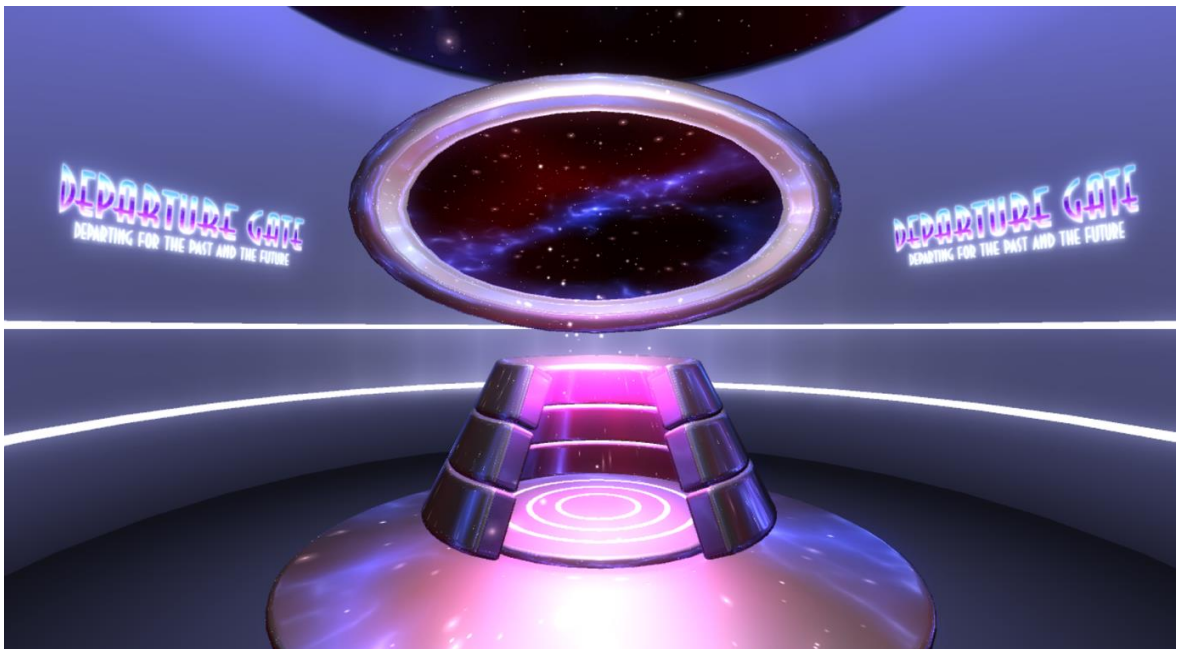
**Figure 4. Introduction to Moneyworld, Time machine chamber.**

After the time travelling, the participant is transferred to a corner store in the 1960s were the main interaction takes place. The virtual shop designed in this research is based on a typical 1960s corner shop with the items displayed behind the counter. Figure 5 shows the shop-keeper in the corner shop. The interaction starts with a tutorial by the same unembodied voice, introducing the old money system to the participant. After the review, the voice demonstrates how to use the coins in order to buy items.

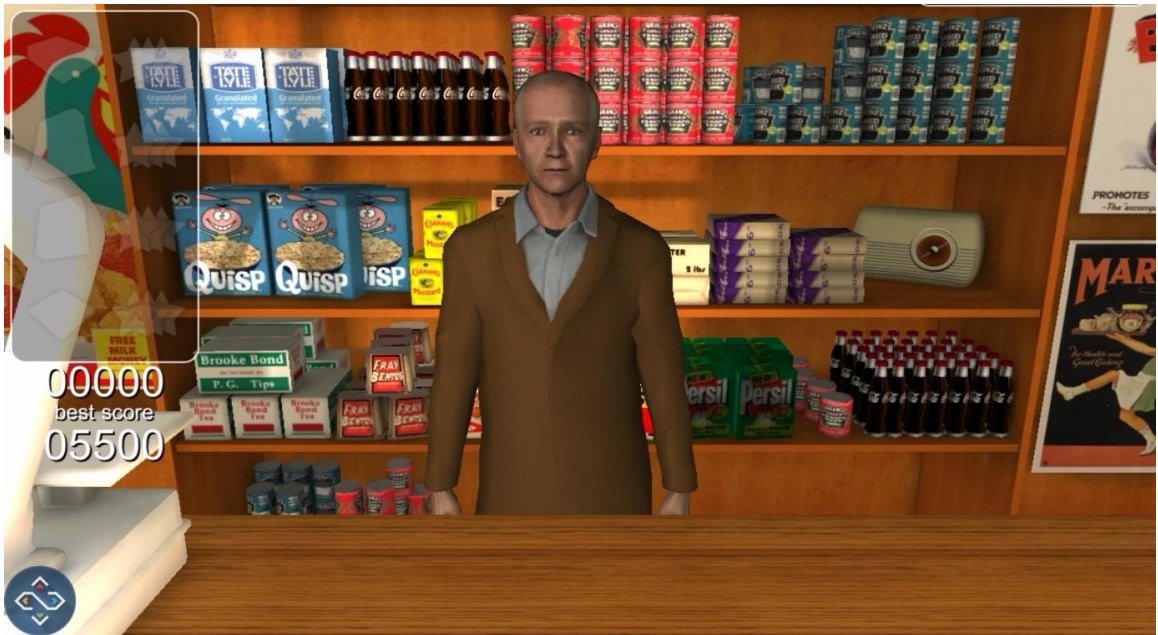

**Figure 5. Corner store layout with shopkeeper ECA.**

After the introduction the application starts with a small tutorial on the gameplay delivered by another agent, Alex (instructor). Alex provides background information to the user on the currency and assistance when needed. In the 1960s, the currency used in Britain was an old monetary system based on pounds and shillings which was denominated by 12.  After the description, Alex asks the participant to review the coins via an understanding exercise through speech. Associated error
 recovery dialogue was included for instances where the user was silent or answered with an incorrect response.  Alex also tells the user which item to purchase in the shop.  Figure 6 depicts Alex in the virtual portal within the shop.

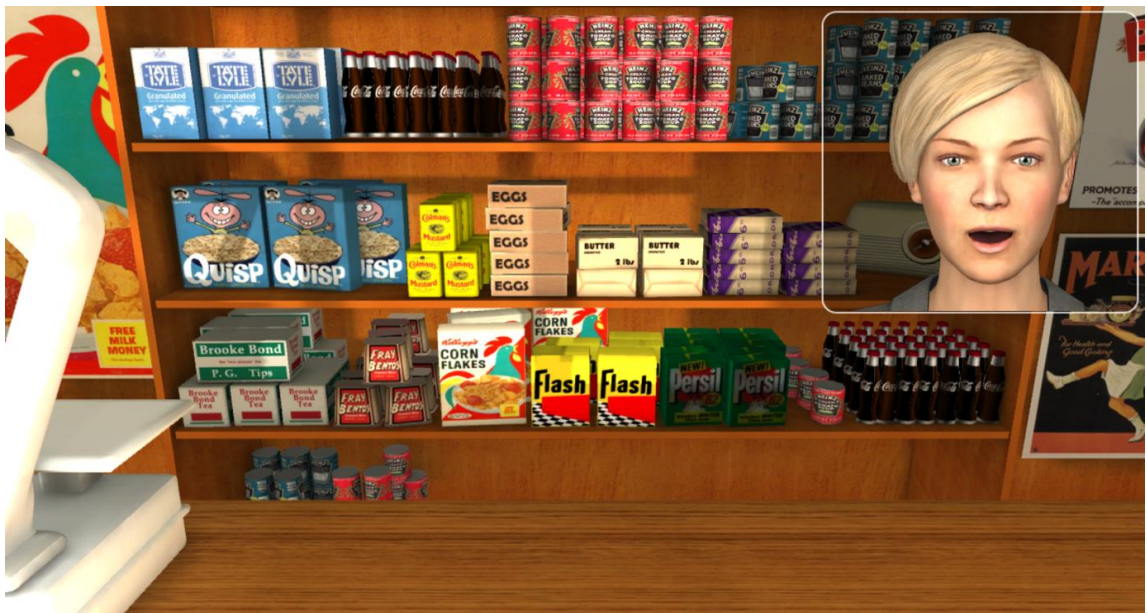

**Figure 6. Alex shown in the virtual portal**

After Alex's tutorial, she introduces the multimodality of the application, that of the coin submission tray. For the user t pay for the products in the shop, a virtual wallet is presented on the bottom of the screen with all the coins (see Figure 7).

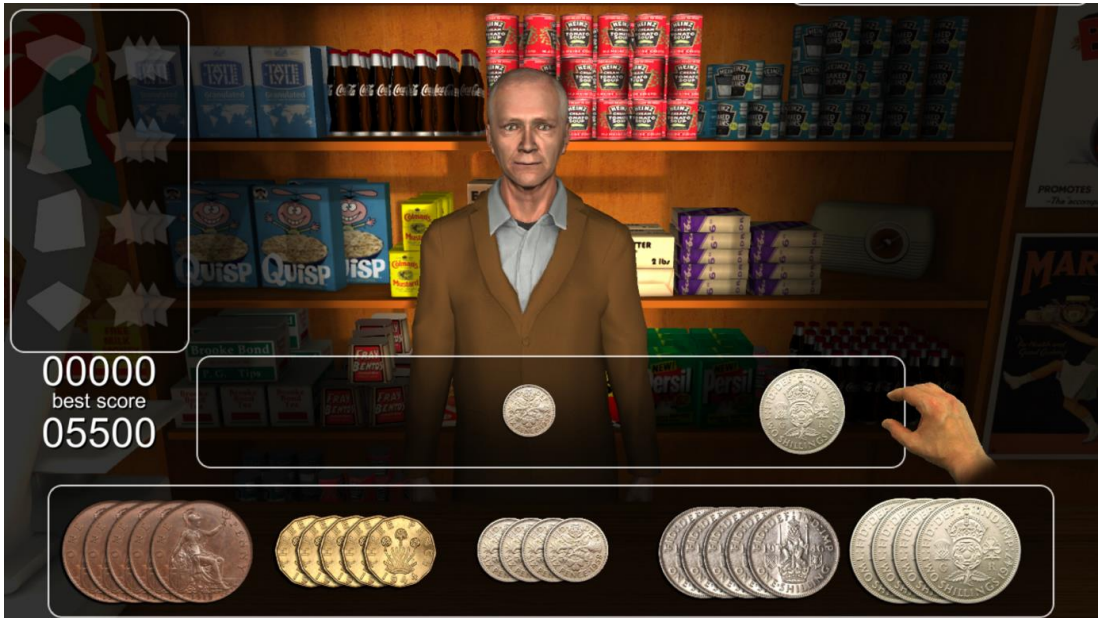

**Figure 7. Coin submission.**

In the neutral text version of Moneyworld, both the instructor agent (Alex) and the collaborator agent (shopkeeper) were presented in the form of a neutral text, as shown in Figure 8.

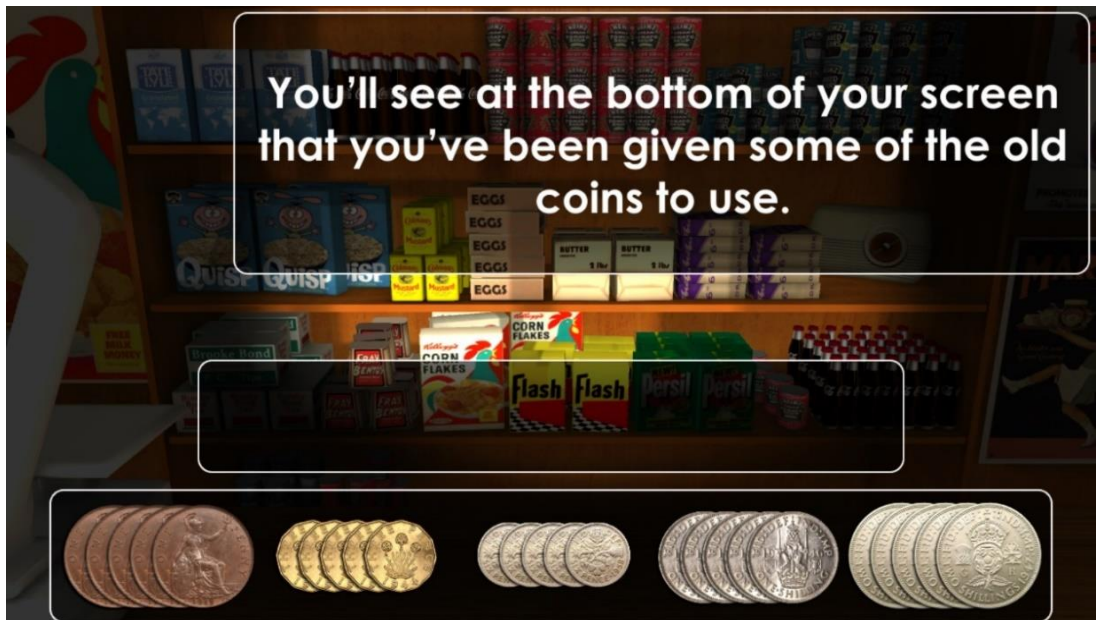

**Figure 8. Neutral text instructor.**

In total, the user is given four items on their shopping list to 'buy' at the virtual shop and is given feedback after each item for correct payment made, efficiency of payment (payment made with the fewest number of coins), and efficiency of task (whether any additional help was required for each item on the shopping list).

## 4. AN EVALUATION OF SPOKEN HECAS WITHIN MSGS

This experiment investigates user attitudes to two versions of Moneyworld involving speech recognition and conversational agents. The objective of this experiment is to examine the extent to which the illusion of humanness evoked by a conversational agent affects the usability of the application and the users' attitudes towards agents with different roles.

**R1:** To what extent do HECAs affect the usability of a mobile serious game?

**R2:** To what extent do users perceive a difference in agent persona between ECA and neutral text presentation as measured by the agent persona instrument (API)?

**R3:** Which factors relating to the HECA's persona attributes account for variability in usability, and to what extent?

### 4.1. Experimental Design

A 2x2 factorial repeated measures experimental design was adopted for this study as the application had two different factors each constituted by two levels as it is shown in Table 2.

The columns of the table represent the two shopping lists used to avoid overexposure between designs, and the rows represent the level of humanness of the agents used (text-low humanness level, HECA-high humanness level). There was no hypothesis for the shopping lists.

| 2x2 design | | Shopping list | |
|---|---|---|---|
| | | 1 | 2 |
| Agent | HECA | V1 | V3 |
| | Text | V2 | V4 |

**Table 2. 2x2 factorial design table**

A power calculation was conducted with G*Power to determine the required sample size. To conduct a two-tailed t-test in order to detect a small effect (d= 0.3) (Cohen, 1988), with an alpha value of 0.05, and a power of 0.8 and a repeated measures design, 90 participants were required.

| Title | | Usability Evaluation: Presence of Humanoid Animated Agents on Mobile Serious Game |
|---|---|---|
| Design | | Repeated measures |
| Null Hypothesis | | There is no difference in usability ratings between software version |
| | | There is no difference in API ratings between software version |
| Dependent Variables | | Usability Questionnaire Responses (1-7 Likert scale) |
| | | Agent Persona Instrument (1-5 Likert Scale) |
| Other Data | | Exit Interview Answers |
| (Experiment) Independent Variables: | 1 | Agent Embodiment (2 levels) |
| Other Variables: | Presentation Order | Agent presentation order randomised. |
| Other Variables: | Shopping list Order | Shopping list presentation order randomised. |
| | Researcher Differences | Controlled by following a prepared procedure and script. |
| | Location | Social space in a university building |
| Cohort | | N = 90 |
| | | power.t.test (power=0.8, d=0.3, sig.level=0.05, type="paired") |
| Remuneration | | £10 |
| Duration: | | 45-60 minutes |

**Table 3. Summary Table of Usability Evaluation: Presence of Humanoid Animated Agents in Mobile Serious Game.**

### 4.1.1. Participants

A total of 90 participants were recruited for this experiment, with an age of under 40 years old. The age limit was calculated based on the context of the game, since the old sterling coins that were used for the game were in circulation till 15 February 1971. Therefore, it was highly unlikely for someone under 40 years old to have knowledge about the old money system. The participants were balanced for version and shopping-list order.

Data were collected from a cohort of 90 participants (47 males, 43 females) with an average age of 25.6 years old. Most participants were international students and professionals (38 native language English, 7 Chinese, 13 Greek, 3 Russian-Ukrainian, 1 Bulgarian, 2 French, 2 German, 3 Hindi, 3 Italian, 1 Indonesian, 1 Japanese, 2 Lithuanian, 3 Romanian, 6 Spanish, 1 Malay, 1 Polish, 1 Telugu, 1 Palestinian Arabic; some were bilingual).

### 4.1.2. Materials

For this research, two validated questionnaires were used: one to assess the usability of the application and two identical questionnaires (API), one for each agent.

The usability questionnaire used in this evaluation is a standardised and validated metric for assessing usability (Jack et al., 1993, Doolin, 2013). Previous research (Dutton, et al., 1993; Jack, et al., 1993; Love, et al., 1992) has identified salient attributes of the perceived usability of interactive systems. The result of this research is the CCIR MINERVA usability questionnaire that was chosen for this research which has been developed and tested as a tool for assessing users' attitudes (McBreen, 2002; Gunson, et al., 2011). The validity of the questionnaire was confirmed by experimental work (Jack et al., 1993).

| Usability Questionnaire Statements |
|---|
| 1. I found Moneyworld confusing to use |
| 2. I had to concentrate hard to use Moneyworld |
| 3. I felt flustered when using Moneyworld |
| 4. I felt under stress when using Moneyworld |
| 5. I felt relaxed when using Moneyworld |
| 6. I felt nervous when using Moneyworld |
| 7. I found Moneyworld frustrating to use |
| 8. I felt embarrassed while using Moneyworld |
| 9. While I was using Moneyworld I always knew what I was expected to do |
| 10. I felt in control while using Moneyworld |
| 11. I would be happy to use Moneyworld again |
| 12. I felt Moneyworld needs a lot of improvement |
| 13. I enjoyed using Moneyworld |
| 14. I thought Moneyworld was fun |

| Usability Questionnaire Statements |
| --- |
| 15. I felt part of Moneyworld |
| 16. I found the use of Moneyworld stimulating |
| 17. Moneyworld was easy to use |
| 18. I liked the voices in Moneyworld. |
| 19. I thought the voices in Moneyworld were very clear. |
| 20. I thought Moneyworld was too complicated |

**Table 4. Usability attributes.**

The second questionnaire that has been used for this research was also a validated metric for assessing the agent's persona called agent persona instrument (API) (Baylor and Ryu, 2003) as shown in Figure 9.

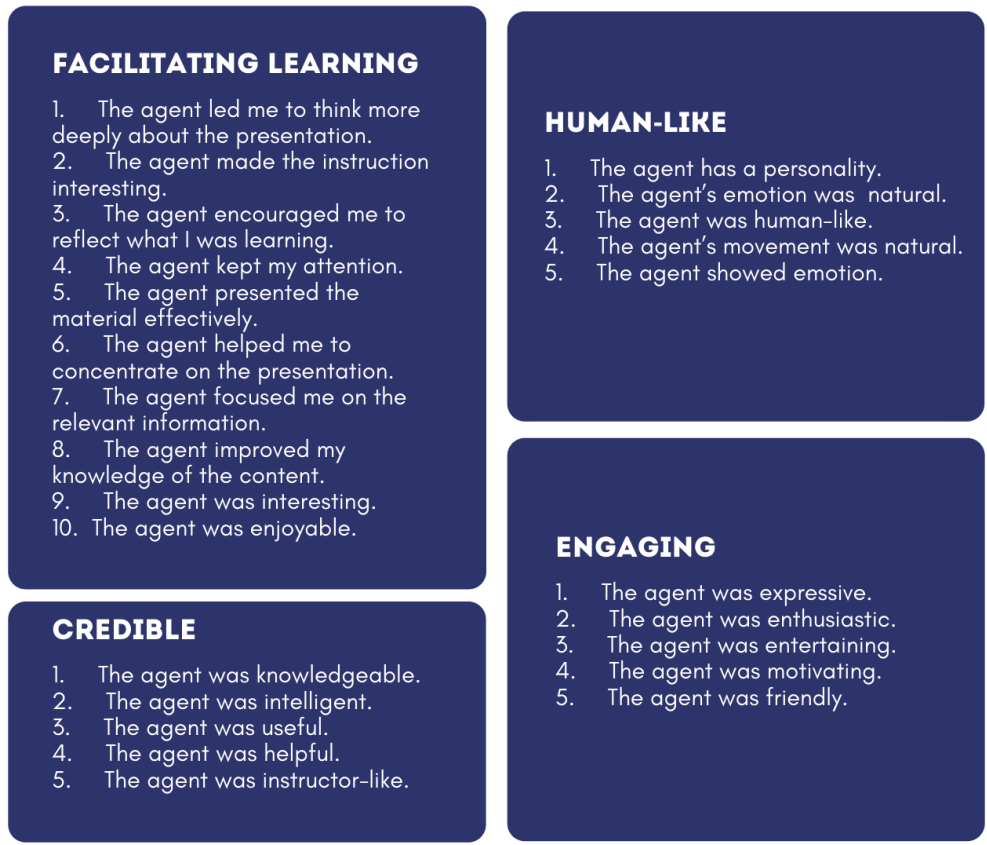

**Figure 9. The API (Agent Persona Instrument) attributes (Baylor and Ryu, 2003).**

The questionnaires were modified to fit the context of the application; therefore, irrelevant Likert items were removed, namely the item "The agent's movement was natural".

Responses for the usability questionnaire were on a Likert-type scale, ranging from 1 = "Strongly agree", 2 = "Agree", 3 = "Slightly agree", 4 = "Neutral", 5 = "Slightly disagree", 6 = "Disagree", 7 = "Strongly disagree". Responses for the API were on a Likert-type scale, ranging from 1 = "Strongly disagree", 2 = "Disagree", 3 = "Neutral", 4 = "Agree", 5 = "Strongly agree".

An exit interview was designed in order to retrieve information on the following topics:

- Participant's view of the use of spoken HECAs and text conversational agent in a mobile serious game.

- The effective deployment of spoken HECAs and text conversational agents in the interface.

### 4.2. Experimental Procedure

The experiment took place in open space workstation within a university building communal space. The setup allowed for observation under circumstances where ambient noise and other people are present. This was in order to simulate the conditions in which people might use ECAs on mobile devices for real applications.

The device used was a smartphone with the following specifications: Super AMOLED capacitive touchscreen, 16M colors, 5.1 inches, 71.5 cm2 (~70.7% screen-to-body ratio), Resolution of 1440 x 2560 pixels, 16:9 ratio (~577 ppi density), Android 5.0.2 (Lollipop), upgradable to Android 8.0 (Oreo) OS and Exynos 7420 Octa (14 nm) chipset.

First, the participants were informed about the purpose of the experiment and then they started the tutorial. In the tutorial a female unembodied voice welcomed the participants and introduced the concept of the game. The user went through the teleporter and the time/space channel and arrived at the 1960s corner shop in order to play the game. In the corner store, the same voice introduced the old coins to the participant followed by a coin review dialogue. The same voice then asked the user to identify three coins from the set and to state the value of each of them in pence. After the review, the voice demonstrated how to use the coins in order to buy items.

The tutorial was the same for both versions and was experienced once at the beginning of the session. A different voice than that of Alex, the assistant/instructor, was used for the tutorial in order to avoid overexposure of one style over the other. After each participant interacted with the tutorial, they were asked to answer some relevant questions to the tutorial.

After finishing with the tutorial's questionnaire, the user played Version 1 of Money World, where they were asked to buy 4 items by Alex, the assistant/instructor, who appeared on the right-top corner window, followed by Version 2. The scene comprised the corner store; the shopkeeper/collaborator that the player interacted with in order to buy items as dictated by Alex; and on the left side there was an inventory of the items purchased and the rewards system.

### 4.3. Analysis Method

Research question one was answered by a paired t-test analysis on the Usability questionnaire data; research question two was answered by paired t-test analysis on the API questionnaire data. An overall score for each questionnaire was calculated as the mean of its items, and then paired t-tests scores and Cohen's d effect size were computed. Subsequent paired t-tests and effect size calculations were conducted on each questionnaire item, with Bonferroni Correction and Holm-Bonferroni Sequential Correction to correct for Type 1 errors. Research question 3 is answered by performing a multiple regression analysis with data from both the usability and the API questionnaires.

Multiple linear regression analysis estimates the coefficients of a linear equation, involving multiple independent variables (IVs), that best predict the value of the dependent variable (DV). In this research, there was no prior knowledge to select some variables based on previous research as no prior research looked on the relationship of the agent's persona and usability. The predictors used in this research were informed by the nature and theoretical base of the experiment.

From the literature (Tibshirani & Hastie, 2016), it is known that sparser statistical models perform better and tackle the problem of overfitting. Thus, a reduction of complexity was achieved by selecting the IVs from theory rather than using all 24 predictors. Another reason for not using all 24 items as IVs is for model interpretability; by removing irrelevant features a model is more easily interpreted. The data was first assessed for normality using visual representations, tests for skewness and kurtosis and z-scores.

Since this research focuses mostly on the affective effect of the HECA using the API instrument, the variables selected for the model belong to the "Emotive interaction" latent variable; this variable is subdivided into the "Human-like" factor and the "Engaging" factor. According to Baylor (Baylor and Ryu, 2003) who developed the instrument: "The characteristics of the Engaging factor represent the social richness of the communication channels (Whitelock et al., 2000) and play an important role to provide 'personality' to the agent and enhance the learning experience", while "the Human-like factor of pedagogical agent persona is what makes it figuratively 'real'. Thus, both the Human-like factor and Engaging factors shape the pedagogical agent's social presence and personality". That limits the number of predictors to 9 ("The agent was human-like", "The agent was entertaining", "The agent was friendly", "The agent has a personality", "The agent showed emotion", "The agent emotion was natural", "The agent was enthusiastic", "The agent was expressive" and "The agent was motivating").

An *a priori* sample size calculation for multiple regression was performed. Based on the rule of thumb that 10 to 15 samples are needed per predictor, 90 samples for 9 predictors should suffice (Tabachnick and Fidell, 2001).

In this research, the ordinary least squares (OLS) full model is used with 9 items as predictors and the usability mean value for the shopkeeper agent. The method used is the hierarchical multiple linear regression, since from theory the "Human-like" factor is more relevant (Model 1: 4 predictors) and is followed by the "Engaging" factor (5 predictors). Model two is a combination of the "Human-like" and "Engaging" attributes and includes the following variables: "The agent was human-like", "The agent was entertaining", "The agent was friendly", "The agent has a personality", "The agent showed emotion", "The agent emotion was natural", "The agent was enthusiastic", "The agent was expressive" and "The agent was motivating".

One case was deemed to be an outlier. The outlier was not removed, instead the mean score for the ECA version was corrected with the next highest score plus one unit as suggested by Field (2013).

## 4.4. Research Question 1: Usability Questionnaire Results

An overall mean usability score was calculated from the 18 usability attributes scores for each of the two treatment groups. The overall mean scores for the questionnaire taken differed between the two versions. The ECA version received the highest overall mean score of 5.32 (which translates to slightly agree on overall usability), while the Text version received a score of 4.40 (which translates to Neutral on overall usability). Table 5 and Figure 10 detail the descriptive statistics for the mean scores of the two versions.

| | Descriptive Statistics | | | |
|---|---|---|---|---|
| | Order of experience | Mean | Std. Deviation | N |
| ECA MEAN | ECA first | 5.21 | .72 | 45 |
| | Text first | 5.43 | .80 | 45 |
| | Total | 5.32 | .76 | 90 |
| TEXT MEAN | ECA first | 4.20 | 1.06 | 45 |
| | Text first | 4.60 | .95 | 45 |
| | Total | 4.40 | 1.02 | 90 |

**Table 5. Descriptive statistics of usability questionnaire**

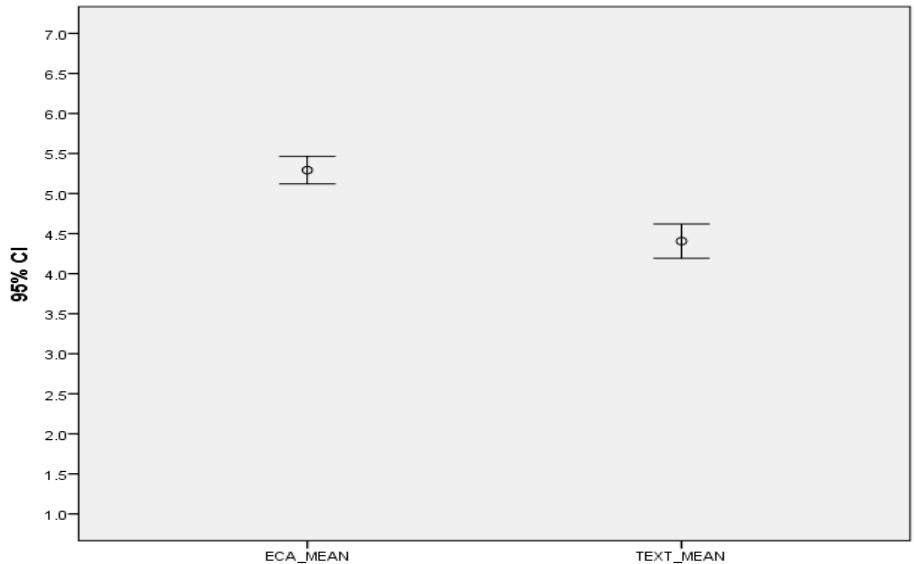
**Figure 10. Summary of usability questionnaire.**

H₀: There is not a statistically significant difference between HECA mean and Text mean.

Hₐ: There is a statistically significant difference between HECA mean and Text mean.

The assumption of normality was examined using a one-sample Kolmogorov-Smirnov (KS) test and both were normally distributed. A dependent sample *t*-test for paired means showed a statistically significant difference between the two mean scores (t=9.45; df=89; p.=0.000) and therefore the null hypothesis can be rejected. Based on the data from the paired samples t-test summarised in Table 6, the null hypothesis can be rejected for all the individual items in the questionnaire.

**Paired samples test***

| | | Mean ECA version | Mean Text version | T | df | Sig. (2-tailed) | Std. Deviation |
|---|---|---|---|---|---|---|---|
| Pair 1 | ECA1 - TEXT1-I found Moneyworld confusing to use. | 5.93 | 5.01 | 5.485 | 89 | .000 | 1.60 |
| Pair 2 | ECA2 - TEXT2-I had to concentrate hard to use Moneyworld. | 5.18 | 4.28 | 5.135 | 89 | .000 | 1.66 |
| Pair 3 | ECA3 - TEXT3-I felt flustered when using Moneyworld. | 5.44 | 4.42 | 6.169 | 89 | .000 | 1.57 |
| Pair 4 | ECA4 - TEXT4-I felt under stress when using Moneyworld. | 5.83 | 4.76 | 6.618 | 89 | .000 | 1.55 |
| Pair 5 | ECA5 - TEXT5-I thought Moneyworld was too complicated. | 6.16 | 5.84 | 3.209 | 89 | .002 | 0.92 |

|  |  | Mean ECA version | Mean Text version | T | df | Sig. (2-tailed) | Std. Deviation |
|---|---|---|---|---|---|---|---|
| Pair 6 | ECA6 - TEXT6-I felt nervous when using Moneyworld. | 5.49 | 4.92 | 3.567 | 89 | .001 | 1.51 |
| Pair 7 | ECA7 - TEXT7-I found Moneyworld frustrating to use. | 5.29 | 3.77 | 7.913 | 89 | .000 | 1.83 |
| Pair 8 | ECA8 - TEXT8-I felt embarrassed while using Moneyworld. | 5.32 | 4.70 | 3.433 | 89 | .001 | 1.72 |
| Pair 9 | ECA9 - TEXT9-I felt Moneyworld needs a lot of improvement. | 4.17 | 2.99 | 7.747 | 89 | .000 | 1.44 |
| Pair 10 | ECA10 - TEXT10-I felt in control while using Moneyworld. | 5.20 | 4.23 | 5.356 | 89 | .000 | 1.71 |
| Pair 11 | ECA11 - TEXT11-I would be happy to use Moneyworld again. | 5.18 | 4.24 | 5.913 | 89 | .000 | 1.50 |
| Pair 12 | ECA12 - TEXT12-I felt I relaxed when using Moneyworld. | 5.04 | 4.32 | 4.481 | 89 | .000 | 1.53 |
| Pair 13 | ECA13 - TEXT13-I enjoyed using Moneyworld. | 5.26 | 4.30 | 5.433 | 89 | .000 | 1.67 |
| Pair 14 | ECA14 -TEXT14-I thought Moneyworld was fun. | 5.22 | 4.30 | 6.144 | 89 | .000 | 1.42 |
| Pair 15 | ECA15 -TEXT15-I felt part of Moneyworld. | 4.64 | 3.51 | 7.060 | 89 | .000 | 1.52 |
| Pair 16 | ECA16 - TEXT16-I found the use of Moneyworld stimulating. | 4.76 | 4.26 | 3.554 | 89 | .001 | 1.34 |
| Pair 17 | ECA17 - TEXT17-Moneyworld was easy to use. | 5.81 | 4.98 | 4.891 | 89 | .000 | 1.62 |
| Pair 18 | ECA18 - TEXT18-While I was using Moneyworld I always knew what I was expected to do. | 5.34 | 4.47 | 3.990 | 89 | .000 | 2.09 |

*99.9972% Confidence Interval of the Difference alpha=0.0027*

**Table 6. Sample t-test summary after Bonferroni correction.**

There was a significant difference in all individual usability items. The ECA version scored higher on all questions. This difference support that the illusion of humanness effect theory holds in participants' perceptions of the software usability. The Text version scored below neutral in 3 attributes (frustration, needs a lot of improvement and immersion) and over slightly agree in only 2 (confusing to use and too complicated). The ECA version scored overall above average and was perceived to be usable. It scored between neutral and slightly agree in 3 attributes (needs improvement, stimulation and immersion), and over agree in all the rest except one where it was scored as strongly agree; that translates to participants feeling that the version was not too complicated.

## 4.5. Research question 2: Agent Persona Instrument Analysis

In this experiment there were two agents, an instructor agent (Alex) that gives instructions on how the coins should be used and says which items should be purchased next and a collaborator agent which interacts with the user during the transaction (the shopkeeper). Although there were two agents in each version, they were assessed and analysed separately as they serve different purposes in the interaction and they are different on many levels; therefore, the two agents cannot be aggregated.

The overall mean scores for the collaborator agent questionnaire differed between the two versions. The ECA agent received the highest overall mean score of 3.67 which translates to between neutral and agree and that participants reacted positively to the agent. The Text agent received a score of 2.81 which translates to between disagree and neutral about their reaction towards the agent. Table 7 details the descriptive statistics for the mean scores of the two versions.

|  |  | Mean | Std. Deviation |
|---|---|---|---|
| Collaborator Mean ECA version | Total | 3.67 | .58 |
| Collaborator Mean TEXT version | Total | 2.81 | .69 |
| Instructor agent ECA Mean |  | 3.54 | .58 |
| Instructor agent TEXT Mean |  | 2.91 | .68 |

**Table 7. Descriptive statistics for the API**

The overall mean scores for the instructor agent questionnaire taken differed between the two versions. The ECA version received the highest overall mean score of 3.54 which translates to between neutral and agree and, thus, participants reacted positively to the agent. The Text version received a score of 2.91 which translates to between disagree and neutral on their reaction towards the agent. Table 8 details the descriptive statistics for the mean scores of the two versions on individual items.

The assumption of normality was examined using a one-sample Kolmogorov-Smirnov (KS) test and both were normally distributed for both agents. There was a statistically significant difference between the two mean scores for the collaborator agent (t=13.068; df=89; p.=0.000; d = 1.34); therefore, the null hypothesis was rejected. In addition, there is a statistically significant difference between the two mean scores for the instructor (t=8.428; df=89; p.=0.000; d= 1.34); therefore, it is assumed that there is a statistically significant difference between the ECA mean and Text mean for the instructor-agent persona questionnaire. The effect sizes are considered large by Cohen's heuristic (Cohen, 1992).

| Questionnaire statement | ECA (Mean =) | TEXT (Mean =) | t | df | p. | Std. Deviation |
|---|---|---|---|---|---|---|
| The agent kept my attention. - | 4.01 | 3.28 | 5.86 | 89 | .000 | 1.19 |
| The agent made the instruction interesting. - | 3.79 | 2.52 | 12.11 | 89 | .000 | 0.99 |
| The agent presented the material effectively. - | 4.09 | 3.64 | 3.69 | 89 | .000 | 1.14 |
| The agent helped me to concentrate on the presentation. - | 3.73 | 3.12 | 5.01 | 89 | .000 | 1.16 |
| The agent was knowledgeable. - | 3.68 | 3.21 | 4.35 | 89 | .000 | 1.02 |
| The agent encouraged me to reflect what I was learning. - | 3.23 | 3.00 | 1.83 | 89 | .070 | 1.21 |
| The agent was enthusiastic. - | 3.68 | 2.29 | 10.93 | 89 | .000 | 1.21 |
| The agent led me to think more deeply about the presentation. - | 3.16 | 2.68 | 4.18 | 89 | .000 | 1.08 |
| The agent focused me on the relevant information. - | 3.69 | 3.61 | 0.69 | 89 | .493 | 1.07 |
| The agent improved my knowledge of the content. - | 3.50 | 3.26 | 1.87 | 89 | .065 | 1.24 |
| The agent was interesting. - | 3.72 | 2.50 | 11.97 | 89 | .000 | 0.97 |
| The agent was enjoyable. - | 3.78 | 2.50 | 11.32 | 89 | .000 | 1.07 |
| The agent was instructor-like. - | 2.80 | 3.53 | -5.16 | 89 | .000 | 1.35 |
| The agent was helpful. - | 3.86 | 3.53 | 2.86 | 89 | .005 | 1.07 |
| The agent was useful. - | 3.82 | 3.59 | 1.94 | 89 | .056 | 1.14 |
| The agent showed emotion. - | 3.76 | 1.81 | 17.88 | 89 | .000 | 1.03 |
| The agent has a personality. - | 3.96 | 1.94 | 18.87 | 89 | .000 | 1.01 |
| The agent's emotion was natural. - | 3.29 | 2.53 | 4.87 | 89 | .000 | 1.47 |
| The agent was human-like. - | 3.78 | 2.06 | 13.73 | 89 | .000 | 1.19 |
| The agent was expressive. - | 3.81 | 2.12 | 12.63 | 89 | .000 | 1.27 |
| The agent was entertaining. - | 3.77 | 2.23 | 12.51 | 89 | .000 | 1.16 |
| The agent was intelligent. - | 3.34 | 2.86 | 5.07 | 89 | .000 | 0.92 |
| The agent was motivating. - | 3.53 | 2.69 | 7.97 | 89 | .000 | 1.01 |
| The agent was friendly. - | 4.34 | 3.06 | 11.87 | 89 | .000 | 1.03 |

**Table 8. Mean scores and results of paired t- tests on Individual Agent Persona Instrument for version - Collaborator agent.**

The HECA version of the collaborator scored higher than the text version on all cases but one (The agent was instructor-like).

As seen in Table 9, the Text agent scored below neutral in 11 attributes (made the instruction interesting, enthusiastic, made me think more deeply about the presentation, interesting, enjoyable, natural emotion, human-like, expressive, entertaining, intelligent, motivating, friendly) and over agree in only 1 (kept my attention). ECA agent scored above neutral in all attributes apart from 1 (the agent is instruction like). It scored over agree in 1 attribute (kept my attention). The difference in the overall mean API scores of the two versions of the game could be attributed to all the items. Based on the literature review, this research focused more on the items of the Human-Like factor of the questionnaire (highlighted in orange) while 6 additional attributes (the ECA: made the instruction interesting, was not instructor-like, was expressive, was entertaining, was friendly and was human-like) were found to have the biggest difference.

| Questionnaire statement | ECA (Mean =) | TEXT (Mean =) | t | df | p. | Std. Deviation |
|---|---|---|---|---|---|---|
| The agent kept my attention. - | 3.87 | 3.08 | 6.01 | 89 | .000 | 1.23 |
| The agent made the instruction interesting. - | 3.48 | 2.63 | 6.40 | 89 | .000 | 1.25 |
| The agent presented the material effectively. - | 4.17 | 3.59 | 5.01 | 89 | .000 | 1.08 |
| The agent helped me to concentrate on the presentation. - | 3.84 | 3.24 | 4.83 | 89 | .000 | 1.18 |
| The agent was knowledgeable. - | 3.96 | 3.44 | 5.70 | 89 | .000 | 0.85 |
| The agent encouraged me to reflect what I was learning. - | 3.49 | 2.99 | 3.75 | 89 | .000 | 1.27 |
| The agent was enthusiastic. - | 3.10 | 2.44 | 4.80 | 89 | .000 | 1.30 |
| The agent led me to think more deeply about the presentation. – | 3.27 | 2.73 | 4.40 | 89 | .000 | 1.15 |
| The agent focused me on the relevant information. | 4.09 | 3.70 | 4.06 | 89 | .000 | 0.91 |
| The agent improved my knowledge of the content. | 3.83 | 3.34 | 3.47 | 89 | .001 | 1.33 |
| The agent was interesting. - | 3.24 | 2.61 | 5.15 | 89 | .000 | 1.17 |
| The agent was enjoyable. - | 3.29 | 2.62 | 6.02 | 89 | .000 | 1.05 |
| The agent was instructor-like. - | 4.27 | 3.80 | 3.90 | 89 | .000 | 1.13 |
| The agent was helpful. - | 4.12 | 3.67 | 4.67 | 89 | .000 | 0.93 |
| The agent was useful. - | 4.00 | 3.79 | 2.07 | 89 | .041 | 0.97 |
| The agent showed emotion. - | 2.92 | 2.01 | 7.95 | 89 | .000 | 1.09 |
| The agent has a personality. - | 3.10 | 2.11 | 7.67 | 89 | .000 | 1.22 |
| The agent's emotion was natural. - | 2.99 | 2.52 | 3.40 | 89 | .000 | 1.30 |
| The agent was human-like. - | 3.20 | 2.22 | 7.30 | 89 | .000 | 1.27 |
| The agent was expressive. - | 3.16 | 2.16 | 8.04 | 89 | .000 | 1.18 |
| The agent was entertaining. - | 2.89 | 2.46 | 3.43 | 89 | .001 | 1.20 |

| Questionnaire statement | ECA (Mean =) | TEXT (Mean =) | t | df | p. | Std. Deviation |
|---|---|---|---|---|---|---|
| The agent was intelligent. - | 3.54 | 2.94 | 6.09 | 89 | .000 | 0.93 |
| The agent was motivating. - | 3.44 | 2.77 | 5.84 | 89 | .000 | 1.10 |
| The agent was friendly. - | 3.72 | 3.00 | 5.85 | 89 | .000 | 1.17 |

**Table 9-Mean scores and results of paired t-tests on Individual Agent Persona Instrument for version – Instructor agent.**

All API items were statistically significant. The HECA version of the collaborator scored higher than the text version on all cases. The Text version scored below neutral in 14 attributes (made instruction interesting, encourage to reflect, enthusiastic, think more deeply, interesting, enjoyable, emotional, has personality, natural emotion, human-like, expressive, entertaining, intelligent, motivating, friendly) and over agree in none. The ECA version scored overall above neutral apart from 3 attributes (emotion, natural emotional, entertaining). It scored above agree in 5 attributes (presented the material effectively, focus on the information, helpful, useful, emotive).

### 4.6. Research question 3: Hierarchical Multiple Regression Analysis

#### 4.6.1. Results for the shopkeeper- collaborator agent

The descriptive statistics for the predictors used in the model are presented in Table 10. The skewness and kurtosis for each variable were examined with indices for acceptable limits of ±2 used one predictor variable was skewed. That is a mere indicator of non-normality though, since skewed data often occur due to lower or upper bounds on the data such as Likert data produce (NIST, 2017). Upon further investigation, all the predictors were normally distributed apart from item 24 (The agent was friendly), while the box plots of items 17 and 21 were not balanced but the Stem-Leaf plots, Q-Q plots and histograms indicated a normal distribution. Thus, the data were treated as normal and were analysed parametrically.

|  | Mean | Std. Deviation |
|---|---|---|
| The agent was friendly | 4.34 | .673 |
| The agent was motivating | 3.53 | .837 |
| The agent was entertaining | 3.77 | 1.028 |
| The agent was expressive | 3.81 | .982 |
| The agent was human-like | 3.78 | .957 |
| The agent's emotion was natural | 3.29 | 1.073 |

|  | Mean | Std. Deviation |
|---|---|---|
| The agent has a personality | 3.96 | .873 |
| The agent showed emotion | 3.76 | .916 |
| The agent was enthusiastic | 3.68 | .934 |

**Table 10. Descriptive statistics for agent persona predictors (collaborator agent)**

In a summary, no multivariate outliers existed; the assumption of non-zero variance was met as the predictors vary in value; the assumptions of linearity, homoscedasticity and normality were met; the assumption for independent errors was deemed to be inconclusive; the assumption of multicollinearity has been met; the data were suitably correlated with the dependent variable in order to be examined with multiple linear regression.

For model 1 ("Human-Like" predictors), the strongest and the only statistically significant (p. =0.008) predictor was "The agent was human-like" ($\beta$ = .39). In model 2 (full model with all 9 predictors from both "Human-Like" and "Engaging" factors), two were the most statistically significant predictors, "The agent was human-like" ($\beta$ = .4)(p. = 0.010), "The agent was entertaining" ($\beta$ = .03)(p.=0.05) (see Table 11).

|  | B | SE B | β |
|---|---|---|---|
| **Model 1** |  |  |  |
| **Constant** | 4.06 | 0.38 |  |
| *The agent showed emotion* | -0.13 | 0.12 | -.15 |
| *The agent has a personality* | 0.15 | 0.13 | .18 |
| *The agent's emotion was natural* | -0.01 | 0.09 | -0.02 |
| *The agent was human-like* | 0.31 | 0.11 | .39** |
| **Model 2** |  |  |  |
| **Constant** | 4.08 | .51 |  |
| The agent showed emotion | -0.15 | 0.14 | .18 |
| The agent has a personality | 0.84 | 0.14 | .09 |
| The agent's emotion was natural | -0.02 | 0.9 | -0.03 |
| The agent was human-like | 0.3 | 0.1 | 0.4** |
| The agent was enthusiastic | -0.06 | 0.1 | -0.07 |

|  | B | SE B | β |
|---|---|---|---|
| The agent was expressive | -0.05 | 0.1 | -0.06 |
| The agent was entertaining | 0.2 | 0.1 | 0.3** |
| The agent was motivating | 0.13 | 0.12 | 0.14 |
| The agent was friendly | -0.1 | 0.15 | -0.09 |

*a. Dependent Variable: Usability mean score for embodied conversational agent version*

*Note: \*p < .10, \*\* P< .05, \*\*\* p < .001. n=90*

**Table 11. Hierarchical Multiple Regression Analyses for the Shopkeeper Agent of the Embodied Conversational Agent Version.**

For multiple regression the formula to calculate the effect size is:

$$F^2 = R^2/1 - R^2$$

**Equation 1-Cohen's formula for calculating effect size in multiple regression (Selya, et al., 2012).**

In this case, Cohen's formula gives an effect size $f^2 = 0.297$. This represents a moderate to large effect according to Cohen's guidelines (Cohen, 1988).

For the first model, the 4 independent variables from the "Human-like" factor produced an effect size $R^2$ of .17 ($F(4,85) = 4.28$, $p = .003$) which means that the "Human-like factors" accounted for 17% of the variation in ECA Usability. However, for the final model and all 9 predictors, this value increased to 0.229 ($F(9,80) = 2.64$, $p = .010$) or 23% of the variation in ECA Usability. Therefore, whatever variable entered the model in block 2 and the "Engaging" factors accounted for an extra 6% of the variance. The adjusted $R^2$ shows how well the model can be generalised. It was 0.13 for the first model and 0.142 for the second model which implies that the model with all 9 predictors includes some non-important variables that add noise to the model.

## 4.6.2. Results for the Alex- instructor agent

|  | Mean | Std. Deviation |
|---|---|---|
| ECA_MEAN | 5.3173 | .76617 |
| The agent showed emotion | 2.92 | 1.008 |
| The agent has a personality | 3.10 | .995 |
| The agent's emotion was natural | 2.99 | 1.022 |
| The agent was human-like | 3.20 | 1.041 |
| The agent was enthusiastic | 3.10 | .972 |
| The agent was entertaining | 2.89 | .929 |
| The agent was motivating | 3.44 | .751 |
| The agent was friendly | 3.72 | .750 |
| The agent was expressive | 3.16 | 1.005 |

Table 12. Descriptive statistics for agent persona instrument predictors (Instructor agent)

The relevant assumptions of this analysis were tested prior to the multiple regression analysis. In a summary, no multivariate outliers existed; the assumption of non-zero variance was met as the predictors vary in value; the assumptions of linearity, homoscedasticity and normality were met; the assumption for independent errors has been met; the assumption of multicollinearity has been met; the data were suitably correlated with the dependent variable in order to be examined with multiple linear regression.

For model 1, the strongest predictor that was statistically significant was "The agent was human-like" ($\beta$ = .47). For model 2, the strongest predictor was "The agent was entertaining" (see Table 13).

|  | B | SE B | β |
|---|---|---|---|
| **Model 1** |  |  |  |
| **Constant** | 4.31 | 0.28 |  |
| **The agent showed emotion** | 0.11 | 0.10 | .15 |
| **The agent has a personality** | -0.19 | 0.12 | -.25 |
| **The agent's emotion was natural** | 0.05 | 0.1 | 0.07 |
| **The agent was human-like** | 0.35 | 0.11 | .47** |
|  |  |  |  |

|  | B | SE B | β |
|---|---|---|---|
| **Model 2** | | | |
| Constant | 4.19 | .42 | |
| The agent showed emotion | 0.07 | 0.11 | .09 |
| The agent has a personality | -0.23 | 0.12 | -.30 |
| The agent's emotion was natural | -0.01 | 0.1 | -0.11 |
| The agent was human-like | 0.11 | 0.28 | 0.38** |
| The agent was enthusiastic | 0.08 | 0.11 | 0.11 |
| | | | |
| The agent was expressive | -0.05 | 0.11 | -0.07 |
| The agent was entertaining | 0.30 | 0.11 | 0.36** |
| The agent was motivating | 0.02 | 0.12 | 0.02 |
| The agent was friendly | -0.07 | 0.12 | -0.07 |

a. Dependent Variable: Usability mean score for embodied conversational agent version

Note: *p < .10, ** P< .05, *** p < .001. n=90

**Table 13. Hierarchical Multiple Regression Analyses for Instructor Agent of the Embodied Conversational Agent Version.**

In this case Cohen's formula yields an effect size $f^2 = 0.4$. This represents a large effect according to Cohen's guidelines (Cohen, 1988).

For the first model, the 4 independent variables from the "Human-like" factor produced a $R^2$ of .20 ($F (4,85) = 5.37, p = .001$) which means that the "Human-like factors" accounted for 20% of the variation in ECA version Usability. However, for the final model and all the 9 predictors, this value increased to 0.29 ($F (9,80) = 3.56, p = .00$) or 29% of the variation in ECA version Usability. Therefore, whatever variable entered the model in block 2 and the "Engaging" factors accounted for an extra 9% of the variance. The adjusted $R^2$ was 0.16 for the first model and 0.21 for the second model which implies that not all the predictors contributed to the model significantly.

### 4.7. Qualitative analysis

After interacting with each version, participants were asked to comment on their experience with the application and then specifically on each version. All the answers for the open-ended questions were analysed using thematic analysis (Hayes, 2000).

### 4.7.1. Explicit preference for software version

After experiencing both versions, participants were asked which version of Moneyworld they preferred. They were asked to give their answer in terms of their first or second version experienced, and the answers were re-ordered for each version.

Eighty-one participants (90%) stated that they preferred the ECA version, eight participants (8.9%) stated that they preferred the text version and one participant (1.1%) had no stated preference.

Participants were also asked to give reasons for their answer. The majority of comments about the ECA version mentioned that characters were more fun (20); they preferred interacting with a human/character (16); it was more human-like and natural (14); it was more interactive (27 participants); it was easier (17); and the text version was boring and added cognitive load (15).

Some sample comments made by participants are: "The interaction with humans makes the game engaging."; "The ECAs were more engaging and fun. It was easier for me to understand the instructions."; "The shopkeeper made me feel relaxed. It was more interactive and enjoyable."; "I felt I was very familiar, and it was easy to deal with it. I was interacting with a human, so communication was easy."

For those who preferred the Text version, participants commented that it was clearer (2), reading was faster than the ECA (2) or that they preferred the text version (4). Example comments are: "I prefer to read because it is faster."; "Reading didn't take as long as the ECA version."

### 4.7.2. Agent version

Before asked about the shopkeeper, participants were presented with a laminated picture of the shopkeeper as it appeared on the screen during the shopping task and asked "The interface that you interacted with in order to buy the items on the list looked like this. What did you think about it?". The comments were overall positive. Even though the question did not refer to the agent as "He" but rather asked what they thought about it, most participants commented on the human characteristics of the agent. Some participants (23) thought that the shop-keeper was human-like. Several participants (16) characterised the agent as friendly while others as funny or fun to interact with (26). Some participants (16) commented that they liked him or liked interacting with him and five said that having a person to interact with made the experience better. Example comments made are: "I could imagine how he would be in real life. It was a realistic, human-like character."; "It was human-like. I liked interacting with someone and receiving positive feedback."; "Having human-like characters makes it more captivating and enjoyable."; "He added a personality. It was fun and interesting to interact with him. He gave funny comments."

Seven participants made negative comments on the ECA which mainly had to do with the uncanny valley theory and the face animations but were accompanied by some positive comments like he was fun or friendly e.g. "He was interesting, funny and human-like. He was friendly, but he had a worrying expression."

Participants were asked their opinion on Alex where they were also presented with a picture of the agent as it was presented in the game. Most comments on Alex were positive with 18 participants reporting that they liked her voice and they could focus better due to the voice; nineteen participants identified her role in the interaction as the agent that gave instructions and their perception was positive as they felt Alex was helpful; seven thought she was human-like, while eight stated that the addition of character was better as it made it more natural or easier to focus; thirteen commented that the interaction was more interesting and fun; and twelve that it was more clear. Example comments made are: "More interesting, less boring, human-like."; "Because of the voice I was able to perceive emotion. I think this is a better way to receive instructions."; "She was nice and friendly. She was encouraging and gave clear instructions."; "She was more instructive than the text."

Fourteen comments that were made were on the negative side. Most had to do with the lip synching that was lacking or that she came across as robotic, her face was distracting or was emotionless and that she did not add much to the game. Some examples are: "Her lip synchronisation was not good and this made her funny. She came across as an emotionless robot."; "She was creepy and unnecessary. I do not think that she added anything (any value)."; "It was clear because you get the information. The agent is distracting. It may be boring, but it is clear."

### 4.7.3. Text version

While focusing specifically on the text version, participants were asked their opinion on the agents. Similar to the ECA version, before asked about each agent, participants were presented with laminated pictures of the agents as they were presented in the game.

In the question "The interface you interacted with in order to buy the items on the list looked like this (show text-based shopkeeper). What did you think about it?" only a few comments where positive. Some participants (15) answered that it was clear, straight forward or direct although not human-like or emotional. A few (12) had a lukewarm reaction towards the text shopkeeper by saying that it was fine, good or ok but not engaging. Only six thought that it was easy, five liked it, two thought it was helpful and one said that it helps them focus. Example comments made were: "Good but poor compared to the ECA version which was an improvement."; "It was straightforward, clear but not emotional." However, one participant noted: "It was clear because you get the information. The agent (ECA) is distracting. It may be boring, but it is clear."

Most comments regarding the text version of the shopkeeper were underwhelming and negative. Ten said it was boring and less entertaining. Other comments suggested that they had to concentrate hard to remember the prices and was stressful (14); the text agent was less engaging (eight); it was frustrating to use (eight); and it was confusing (five). A few examples of comments are: "It was stressing for me to read it. It was more difficult to remember the prices."; "I found it easier to understand the task but less engaging, less entertaining and unrealistic."; "It was boring. I did not feel immersed. I was frustrated."; "I got nervous when the text went away because I had to remember. I was not immersed but I was more concentrated on the task. It was a very mechanical experience like an exam." Similar comment were made about the text version of Alex the instructor agent: "It was clear but not interesting."; "It was helpful, but it was not clear that I had to speak."

### 4.7.4. Agent preference

After having been asked about their thoughts on each agent they faced during the game, participants were asked to explicitly state which agent they preferred in each role. Again, participants were presented with screenshots of all the four agents to choose from.

In the question "Which system did you prefer to interact with on the shop?", 76 participants (84.5%) preferred the ECA version, 13 the Text version (14.5%) and one (1%) had no preference. Participants justified selecting the ECA version of the shopkeeper saying they found him more interactive, entertaining, it made the interaction more natural and real, the addition of voice helped them concentrate better and they could focus better. Some of the comments were: "Really liked him. He was polite and funny."; "It felt more like a character, a human."; "It made it seem natural and interactive. I didn't have to focus as much."

Those who preferred the text version of shopkeeper gave comments such as that it was straight forward, quicker and less distracting, such as "The shop keeper was fun, but he was distracting me from understanding and remembering" and "The ECA was slow".

In the question "Which system did you prefer to be assisted from?", 67 preferred the ECA version (74.5%), 18 the text version (20%) and five (5.5%) had no preference. Participants who preferred the ECA version of Alex elaborated on their response by saying that the version with the character was more enjoyable and felt more interactive; it was easier to concentrate and understand the instruction because of the voice; it mimicked human to human interaction and added character; and that unlike the text, the agent made the application feel more like a game.

The participants who preferred the text version of the instructional agent justified their choice with comments such as that having a character did not add to the interaction and it was distracting because the role of the agent was to give instructions. A few of the comments referred to the fact that her facial expressions (ECA Alex) were weird and it was distracting. Also, a few commented that reading instructions was easier or quicker and text was enough for instructions.

### 4.7.5. Use of agents

Finally, participants were asked if they used agent/assistants on their phone and their opinion on speech interfaces and natural language interaction. Again, the answers were organised and analysed for recurring themes.

The first question participants were asked was: "Do you use assistants/agents such as Siri/Cortana/Speaktoit on your smartphone in your everyday life?". The majority (48) stated that they do not use agents on their phone, 30 said that they use agent sometimes, nine answered that they use agent every day and three did not own a smartphone. Out of those who use agents, 22 use Siri, nine Ok Google, four Cortana, two Duolingo, one Google now and one S voice. When asked for what tasks they used agents, 14 answered for fun, 11 for web searching, six for checking the weather, five for calendar and reminders, three for calls, three for setting the alarm, two for texting, two for language learning, two for finding their contacts, two for navigation and two for basic functions.

The next question was "What do you like about this kind of interface?" and "What do you dislike about this kind of interface?". In terms of what participants like, 26 participants responded that speech recognition systems are convenient for hands free situations, 12 said that it is faster than typing, ten answered that it is an easier type of interaction, seven said that it is a fun way to interact and five answered that it is a natural way to interact. Illustrative comments are: "The advantages are that it is hands-free. I find the keys on the phone to be tedious" and "Speaking is faster and more human-like."

To the question "What do you dislike about this kind of interface?", 20 participants responded that speech recognition systems still have issues with picking up accents, 18 answered that using it in public would be embarrassing, 16 said that speech recognition systems need improvement as there are still many voice recognition issues that make the interface frustrating to use and 11 responded that they are used to do things manually. The main concerns for speech systems were privacy and that recognition is not optimal yet. A few of the comments were: "I would be embarrassed in public and I do not want to bother other people" and "Currently it needs improvement as due to accents it is not very reliable."

## 4.8. Summary

This chapter presents the findings of a large-scale evaluation on the effectiveness of spoken HECAs in a mobile serious game.

Results show that perceived usability was statistically significantly higher for the version with the ECAs compared to the neutral text version with a large effect size (Cohen's $d$ = 1.01). The ECA version scored 5.32 while the text version scored 4.40 in a 7 point Likert scale.

When exploring the agents' persona as perceived by the user, data showed that the difference between the ECA and the text version was statistically significant for both agents, with the ECA version scoring higher in both cases. The individual attributes that were the most significant for the shopkeeper/collaborator were: "The agent made the instruction interesting", "The agent was enthusiastic", "The agent showed emotion", "The agent has a personality", "The agent was human-like", "The agent was expressive", "The agent was entertaining" and "The agent was friendly". For the Alexa/instructor agent the most significant attributes were: "The agent showed emotion", "The agent has a personality", "The agent was human-like" and "The agent was expressive".

Upon further analysis, the multiple regression that was conducted in order to identify how much of the variability in usability can be explained by the API attributes, showed that the agents' entertaining, and human-like qualities contributed most to usability for both agents in the scenario.

Qualitative analysis supports the results obtained by the quantitative data with many participants referring to the ECAs as more fun to interact with, more human-like, more engaging, easier to use and making the transaction feel real.

## 5. DISCUSSION

| | |
|---|---|
| **R1:** To what extent do HECAs affect the usability of a mobile serious game (MSG)? | In this study, the use of a HECA resulted in significantly higher usability scores with a large effect size (Cohen's $d = 1.01$). |
| **R2:** To what extent do users perceive a difference in agent persona between ECA and neutral text presentation as measured by the agent persona instrument (API)? | Users' scores on the API were higher when using the ECA version of the software, particularly on items relating to personality, expressive and whether it was human-like. |
| **R3:** Which factors relating to the HECA's persona attributes account for variability in usability, and to what extent? | The agents' entertaining, and human-like qualities contributed most to usability for both agents in the scenario. |

**Table 14.** Summary of answers to research questions.

### 5.1. Support for the illusion of humanness effect

For both agents, the quantitative analysis revealed that the overall mean scores of the API questionnaire did differ between the two versions. The HECA agents received scores of between 3 and 4 (out of 5) on the likert scale which translates to between neutral and agree and that participants reacted positively to the agent. The Text agents received scores of between 2 and 3 which translates to between disagree and neutral about their attitude towards the agent. Further, Cohen's effect size value (d = 1.34) suggested a high practical significance which suggests that the inclusion of an HECA in the role of the collaborator or instructor has a meaningful impact on the API and how participants perceive the agent.

In both agents, that of the collaborator which was the role of the shopkeeper and that of Alex the instructor in this scenario, the same two attributes out of nine were deemed significant for contributing to usability. The first attribute was "The agent was human-like" which is especially important since the underlying theme of the experiment was the illusion of humanness. The variable belongs to the "Human-like" factor which to quote Baylor "address the agent's behaviour and emotional expression in terms of its naturalness and personality." (Baylor & Ryu, 2003). The other factor belonged to the "Engaging" factor, also according to Baylor and Ryu "pertains to the motivational and entertaining features of the agent". Regardless the fact that their role in the interaction was different, for both agents the attributes that contributed more to usability were that they were perceived as human like and as entertaining.

In the case of the shopkeeper the "The agent was friendly.", "The agent showed emotion", "The agent emotion was natural", "The agent was enthusiastic" and "The agent was expressive" variables, even though not significant, had a negative relationship with the DV which can be justified by the uncanny valley theory since the agents' animation and lip-synching weren't flawless thus producing an uncanny feeling, also some comments referred to the shopkeeper as 'overly friendly' and 'creepy'. Feelings like this would detract from the illusion of humanness.

As further support for the illusion of humanness, the qualitative investigation also found comments relating to "human-like" and "entertaining" attributes of the ECA agents. Overall, 55 comments were made for either agents where they were described as human-like or human and 61 comments where they were described as fun and/or entertaining.

The majority of the comments on the Shopkeeper that were positive had to do with the fact that the agent was humanlike (23), made the interaction feel real or referred to as a "real person" (13), he made the interaction fun or he was funny (26) and he was friendly (16). Similar comments were made about the instructor agent where she was described as friendly (18), human-like or like a real person (12) and fun or enjoyable (14). These comments attribute human characteristics or a human dimension to the agent. Further support for the illusion of humanness comes from, participants' use of "he/she" pronouns to refer to the agent when the agent was presented in the ECA form, and researchers' observations that the participants applied social rules and followed similar social cues as in human to human interaction as they waited for the agent to conclude the question before answering. Because of the agent's presence, when the system did not pick up their voice they sympathized with the agent as if he couldn't hear them correctly rather than thinking it was their fault. "I relaxed when the SK said that he did not hear me because it made me feel it was not my fault." And "He was entertaining. The comments made it like it was his fault. He was funny and human like."

## 5.2. Usability considerations for ECAs

The results from the study raise a series of usability issues which researchers and developers could consider when making the choice as to whether include an ECA, or when designing the features of an ECA in a similar context.

### 5.2.1. Concentration and ease of use

The data show that users reported that they had to concentrate harder when using the Text version compared to the ECA version. The empirical data are supported by the qualitative data with 14 participants suggesting that during the text version they had to concentrate hard to remember the prices and was more stressful. This can be connected to the fact that reading from a screen can increase the extraneous cognitive load, while interacting with an ECA did not require to concentrate as hard as there were auditory and visual cues. The explanation is supported by Wik's (2011) previous work who claimed that through task- based interactive exercises with sound, pictures, agents and games, a more robust memory trace is created. The empirical data also support claims by Doumanis (2013) and Van Mulken (1998) that ECAs can improve cognitive functions and that by using ECAs the user can spend their cognitive resources on the primary task. Also, the results contradict one of the main arguments against ECAs, i.e. ECAs can lead to cognitive overload and distract from the main task because participants have to spend cognitive resources in processing visual and auditory information (Walker et al., 1994). The reduced cognitive load compared to the text version contributes to the ECA version by appearing easier to use and demanding less concentration.

### 5.2.2. Frustration and Embarrassment

Users reported feeling more frustrated while using the Text version of the game compared to the ECA version. Participants also commented that they felt the Text version was less responsive. While both versions were identical apart from the control factor, from the observations this can be explained by using the media equation theory. People responded to the questions at the appropriate time when they had visual and auditory cues from the ECA, while on the Text version people responded to the question as soon as they read it (speech input initiated when the question disappeared from the screen for the Text version and when the audio prompt for the question ended for the ECA version) thus making it look non-responsive. The qualitative data support the evidence with eight participants claiming that the text version was frustrating to use and five that it was confusing.

An interesting finding was that although participants reported quite often that they would feel embarrassed using a speech recognition system in public, both versions were rated relatively high although they felt less embarrassed playing the game with the HECA. A possible justification might be the "illusion of humanness" since the unconscious reaction is like that of conversing with a human thus making it less embarrassing.

## 5.3. Fun and enjoyment

Users rated the ECA version as more fun and enjoyable than the Text version. During the exit interview participants commented that the Text version felt outdated, while the ECA version felt more like a game and the graphics resembled more contemporary game. Also, many users commented that the Text version was more neutral, while the shopkeeper's comments and the more human-like interaction made the game more fun. Mulken et al. (1998), while empirically studying the persona effect, found that the presentation was perceived as less difficult and more entertaining even though the presence of an agent had no effect in comprehension. Even though the persona effect focusses more on the effect of agents on learning, the effect of ECAs on entertainment and ease of use is the same as in the empirical work presented here. Another pair of researchers (Koda and Maes,1996) supported that the presence of an ECA in a game application may result in increased entertainment, an assumption that can also be confirmed from the empirical data presented here.

### 5.3.1. Immersion

Especially in game design, immersion is a significant element. Sweetser and Wyeth (2005) list immersion as an element of game flow which is the experience during the act of gaming. The empirical evidence shows that the HECA version scored significantly higher than the Text version in terms of immersion ("I felt part of Money world."). Also, the qualitative data confirmed that participants felt that the ECA version was more immersive and interacted like in a real transaction. This can be justified by the anthropomorphisation of the system and the "illusion of humanness" which mimicked a real-life interaction.

### 5.3.2. Knowing what to do

Users reported feeling like they had a better understanding on what they were expected to do while using the ECA version of the game compared to the Text version ("When I was using Money world, I always knew what I was expected to do"). This can be partially explained by the theory of affordances (perception drives action). While playing the Text version of the game, most participants tried to tap on the items in the background rather than speaking. In the ECA version, due to the visual and auditory cues, they figured out that they had to respond verbally. Since speech interaction is an integral part of this study, the results support that visual and auditory cues evoke a verbal response.

Shneiderman was one of the biggest critics of ECAs. He argued that humanising the system may induce false mental models (Shneiderman and Maes, 1997). An example is that anthropomorphic agents may lead the user to believe that the system is also human-like in terms of cognitive aspects. That can make the user have expectations from the system that it does not possess and may result in a negative experience (Doumanis, 2013) However, in the case of this research participants had the "illusion" that ECAs had human-like cognitive aspects, especially in the case of the shopkeeper, that resulted in a positive experience instead of a negative one. The results support the view of Cassell that: "Humans depend to a great extent on embodied behaviours to make sense and engage in face-to-face conversations. The same happens with machines: embodied agents help to leverage naturalness and users judge the system's understanding to be worse when it does not have a body (Cassell, 2001)."

## 5.4. Does the role of the agent matter?

Although the patterns are broadly similar for both agents, the users' preference data indicates that more participants preferred the ECA version of the shopkeeper agent to the text version than preferred the ECA instructor to text. While 84.5% of participants preferred the HECA version of shopkeeper, the corresponding percentage for ECA Alex was 74.5%. Some of the comments indicate that participants were prone to making comparisons between the two agents even though they recognized that the agents had different roles.

According to participants, Alex's facial expressions were not as responsive as the shopkeeper's. Also, some identified that because this agent gave instructions only they were not bothered by having text. This is because they did not interact with this agent the same way they did with the shopkeeper thus having lower expectations which was further supported by participants' comments. Also, it was observed that when participants experienced the text version first, they preferred the text version of Alex. This was not the case for the shopkeeper agent. The facial animation along with the designated role of the agent as the instructor–with whom they did not interact directly–justifies the larger percentage of participants preferring the text version even though the majority preferred the HECA version. A couple of examples would be: "It was good for instructions, but I did not care much for it" and "It was less interacting, and it was more giving instructions. It was educational."

## 5.5. Limitations, Future work and Implications for Developers

### 5.5.1. Limitations

Even though Moneyworld is a serious game, it was not developed as an educational software. The primary purpose of this evaluation was the usability of the application and how it is affected by the inclusion of HECAs and not learning effectiveness therefore it was not measured. This was a conscious decision as learning is a complex construct making it difficult to measure (Bellotti et al., 2013) while determining whether a serious game is successful at achieving the anticipated learning goals is a time consuming, complex, difficult and expensive process (Hays, 2005; Enfield et al., 2012 ). Chin et al. (2009) attribute part of this difficulty on the fact that video games are inherently open-ended which makes it difficult to collect data.

Even though the advertisement for participants in the main experiment stated clearly that only people with proficient knowledge of English should participate, a few had difficulties in understanding the language in either verbal or textual form. As a result, a small number of participants had to be turned away. Relevant to international participants, a few had a strong accent and the speech recognition system could not easily pick up their voice because it was developed using an English vocal dictionary in Pocket Sphinx. A way to tackle this issue for future experiments would be instead of self- evaluation of English proficiency, prospective participants should complete a test.

A few of the negative comments focused on the ECAs' facial animation. Animating a character by hand is a time consuming and tedious task that not always guarantees a good outcome. For that purpose, there is software that focuses on creating realistic facial and body animation. The main obstacle in the presented research is the financial limitations that did not allow using top tier facial and body animation software which usually costs a few thousand pounds also in equipment and training. That resulted in using software within our budget which created decent animations but there is surely more room for improvement in this area.

At the time of the development of the SG, the ECAs used were built in order to approximate the highest human likeness possible with the means we had. However, there are more design decisions that can be made that can further human likeness most of which are reliant on technological advancements such as those in graphics, CPUs, GPUs, animation etc. Realistic looking characters are already being used in commercial games. It is very promising that we got these results despite the limitations we had, and it would be interesting to replicate this experiment with commercial grade realistic ECAs.

### 5.5.2. Future Work

In the current study the usability questionnaire was developed to measure the participants' subjective impressions of efficiency and effectiveness (system performance). In the future performance could be based on objective scores (effectiveness) and time (efficiency). Further evaluations could identify which aspect of the anthropomorphic interface of ECAs evokes most the illusion of humanness and contributes more to usability. In order to examine that, further evaluations need to be carried out to specify which anthropomorphic elements are the ones evoking an illusion of humanness and affecting usability more (different levels of anthropomorphic agents).

In addition, it is important to replicate this study in other contexts to discover whether the illusion of humanness effect would hold in other software applications. The medium on which Moneyworld was tested was a serious game, but the "illusion of humanness" is not specific to a certain topic or medium and could be used in other contexts. It is possible that users' perceptions of usability of conversational agents vary depending on context – in hands-free, eyes busy situations, users may prefer agents not to be embodied. The illusion of humanness becomes less of an academic issue but more of a real-life issue due to the increasing use of virtual agents and smart screens (virtual agents with a screen) such as Amazon Echo and Google Home in our homes. One possible topic for future evaluation would be testing the addition of ECAs in home smart screens like Amazon Echo show. Would it be worth adding and for which purposes? A Greek company called MLS already has an ECA version incorporated in their smart screen called MAIC but no data on its usability are available.

Participants who took part in this research were in their majority highly educated, with technological literacy and between 18 and 40 years old. It would be worth exploring the illusion of humanness effect on older users or children and people of varying educational and cultural backgrounds as their response to the system might differ.

### 5.5.3. Implications for developers

The development of ECAs is a time-consuming process that developers might not be willing to invest in without evidence showing that it is worth the effort. In application development, assuring usability is an important part of the success of the interaction. In the study reported in this paper, increased usability related to the "illusion of humanness" effect which in turn results from high human likeness.

Due to the methodological approach followed and the attention to the effect sizes, the number of evaluation participants was large enough to allow for a safe generalisation to the population. However, the generalisability of the evaluation findings to the general adult population should be treated with care. When developing usable spoken multimodal systems, the appropriateness of speech interaction must be decided for each application anew based on the purpose and environment of the application (Dybkjær et al. 2004). Weiss (2015) makes a similar claim that whether usability and quality are to be enhanced by using an ECA in a multimodal human-machine interface must be decided for each application anew. Since the platform for this evaluation was a mobile serious game, no generalisation can be made about the "illusion of humanness" in other applications with different purposes or contexts. Nevertheless, the generalisation that can be made safely based on the evaluation findings is that contextually relevant spoken HECAs of high human likeness with collaborative and instructional roles can induce illusion of humanness which results in increased usability in mobile serious games. A suggestion to developers for improving usability in similar contexts would be to incorporate spoken HECAs with high human likeness by following the design decisions in Figure 3. Those decisions are not arbitrary as there is evidence from the literature on what results in high human likeness. Isbister and Doyle (2002) claim that an agent with physical appearance, sound and animation can cause a powerful visceral reaction on the user and evoke the "illusion of life". By enhancing realism in movement, creating natural sounding speech and creating the right visual style that fits the application, user's reaction to the agent can be amplified. Applying however the same ECA design principles by following the ECADM under different circumstances (different media, different game genres, more diverse population etc.) would help determine the extent of the generalisability of the effect.

The ECADM and the spectrum of application interface design in relation to human likeness can be used to inform design decisions on the development of ECAs and the level of human likeness desired respectively. The ECAD model serves a dual function; apart from informing design decisions for designers it can act as a guide to categorise ECA research which will allow for better comparisons and analysis; in ECA research the characteristics of ECAs are not always reported or when they do they lack information that can be used for replication, analysis and comparison.

## 6. CONCLUSIONS

The primary aim of this research was to examine the extent to which spoken humanoid embodied conversational agents (HECAs) affect the usability of a mobile serious game application.

Mixed method analysis allowed for triangulation of findings. Following specific design decisions based on the ECADM model resulted in ECAs with high human likeness. High human likeness in turn resulted in the illusion of humanness effect. The findings suggest that ECAs with high human likeness evoked the illusion of humanness effect and improved the usability of the application.

Results are consistent throughout analyses. The ECA version scored statistically significantly higher than the text version with a large effect size that shows that the results translate to a meaningful real-life difference. The regression analysis showed that the attributes "entertaining" and "human- like" contributed more to usability for both agents which supports the theory that the illusion of humanness has an impact on usability. All quantitative results are supported and further explained by the qualitative data where users used pronouns when referring to the ECAs and justified saying that they were human-like and the interaction was more natural and fun because of them.

In conclusion, ECAs on mobile devices have potential advantages over current interaction paradigms in improving usability because they provide a more "human-like" way of communicating with a complex system.

The implications of these findings are that developers should decide for each application anew if ECAs are fitting to the context and purpose of the application. However, developers should consider that in this context ECAs with high human likeness result in the illusion of humanness which in turn improves the overall usability.